\documentclass[12pt,preprint]{aastex}
\usepackage{graphicx}
\usepackage{epsfig}
\usepackage{multirow}

\shorttitle{Dusty, Accreting Giant Star}
\shortauthors{C. Melis et al.}

\usepackage{color}
\usepackage{graphicx}

\newcommand{\kms}{\mbox{km s$^{-1}$}}

\newcommand{\um}{\mbox{$\mu$m}}

\begin{document}


\title{A Substantial Dust Disk Surrounding an Actively Accreting First-Ascent Giant Star}


\author{C. Melis\altaffilmark{1}, B. Zuckerman} 
\affil{Department of Physics and Astronomy, University of California,
Los Angeles, California 90095-1547, USA}
\email{cmelis@astro.ucla.edu}
\author{Inseok Song}
\affil{Department of Physics and Astronomy, University of Georgia, Athens, GA, 30602-2451, USA}

\and

\author{Joseph H. Rhee, Stanimir Metchev\altaffilmark{2}} 
\affil{Department of Physics and Astronomy, University of California,
Los Angeles, California 90095-1547, USA}


\altaffiltext{1}{Spitzer VGSP Fellow}
\altaffiltext{2}{Current Address: Department of Physics and Astronomy, State University of New York, Stony Brook, NY 11794-3800}


\begin{abstract}
We report identification of the first unambiguous example of what appears to be 
a new class of first-ascent giant stars that are actively accreting gas and dust
and that are surrounded by substantial dusty disks. These old stars, who are 
nearing the end of their lives, are experiencing a rebirth into characteristics 
typically associated with newborn stars. The F2-type first-ascent giant star 
TYC 4144 329 2 is in a wide separation binary system 
with an otherwise normal G8~IV star, TYC 4144 329 1. From Keck near-infrared 
imaging and high-resolution spectroscopy we are able to determine that these two
stars are $\sim$1 Gyr old and reside at a distance of $\sim$550 pc.
One possible explanation for the origin of the accreting material is 
common-envelope interaction with a low-mass stellar or sub-stellar companion.
The gaseous and dusty material around TYC 4144 329 2, as it is similar to the
primordial disks observed around young classical T Tauri stars, could 
potentially give rise to a new generation of planets and/or planetesimals.
\end{abstract}





\section{Introduction}

It is well understood that stars are born in dusty, gaseous cocoons. The 
material enshrouding the star eventually settles into a circumstellar disk
in which planet formation is thought to occur on short timescales. By several 
Myr after the star is born the primordial material
that surrounded it has been accreted by the star, removed from the stellar 
system, or accumulated into larger objects. Thus,
the majority of main sequence and first-ascent giant stars that we see are only
sometimes surrounded by dusty debris-disks 
\citep[e.g.][]{meyer04,chen05,beichman05a,low05,kim05,rhee07,zuckerman95,fekel96,jones08} and 
virtually never surrounded by gaseous disks that are actively accreting onto the
star.

Here we report the first unambiguous example of a dusty, actively accreting
first-ascent giant star, TYC 4144 329 2.
\citet{zuckerman08} discuss the possibility that BP Psc,
a star with bi-polar outflows and an orbiting gaseous ring detected in CO 
emission, may be a first-ascent giant of the same class. However,
lacking a known distance, it is not yet clear if BP Psc is truly a first-ascent
giant star or, rather, a strange, isolated T Tauri star. 
The case of TYC 4144 329 2 is
much more solid, thanks to the existence of an ordinary subgiant
companion, TYC 4144 329 1. In the following sections we discuss our 
observations, determination of the spectral types for both stars, initial
recognition of the infrared excess around TYC 4144 329 2, evidence for accretion
onto the dusty star, and finally a determination of the age of the binary 
system.

\section{Observations}
\label{secobs}





At Mauna Kea Observatory we observed both components of the TYC 4144 329 system 
using the NIRC2 near-infrared imaging camera on the Keck II Telescope and the 
HIRES echelle spectrometer at Keck I (see Table \ref{tabobs}). Our NIRC2 data 
were taken using the natural guide star adaptive optics system 
\citep{wizinowich00} to look for any extended emission around TYC 4144 329 2. 
Due to complications in our data sets from anisoplanatic degradation
we can only place a tentative limit on any extent of TYC 4144 329 2 of 
$\lesssim$70 mas. Future measurements that will compensate for these effects are
planned. Variable adaptive optics corrections prevented flux calibration of the 
TYC 4144 329 images with a flux standard star. To determine the flux  
for TYC 4144 329 2 we assumed TYC 4144 329 1 had photospheric flux values
and computed delta-magnitudes for individual images. The results are
discussed in Section \ref{secirex}.

Spectra of TYC 4144 329 1 and 2 were obtained with the HIRES \citep{vogt94}
spectrometer on two dates (see Table \ref{tabobs}). Spectral data were reduced 
and extracted using the {\it MAKEE} software package. 

Low-resolution grating spectra were obtained for the TYC 4144 329 system
with the KAST Double Spectrograph at the Lick Observatory Shane 3-m 
Telescope. Data obtained on UT 28 June 2008 suffered from
extinction due to smoke from nearby 
fires. Data were reduced and extracted using standard {\it IRAF} tasks. 

\section{Results}

\subsection{TYC 4144 329 Stellar Parameters}
\label{secsptypes}

Our star of interest, TYC 4144 329 2, is part of a visual double star system.
Table \ref{tabpars}
contains information about TYC 4144 329 1, the optical primary star, and
TYC 4144 329 2, the dusty star. We took special care to ensure that we correctly
determined the spectral type of both stars and especially the luminosity class
of TYC 4144 329 1. For both stars we employ line-depth ratios to roughly 
determine absolute temperature and, for TYC 4144 329 1, luminosity. Following
previous works \citep{gray89,strassmeier90,kovtyukh06} we sought out in our
HIRES spectra line pairs with appreciably different excitation potentials for
their lower energy levels. Lines with higher excitation potentials are less
sensitive to fluctuations in temperature while lines with lower excitation 
potentials will show significant line strength variations with small
temperature changes. These line ratio diagnostics have been shown to be 
relatively insensitive to metallicity, rotation, and surface gravity effects
\citep{gray94,kovtyukh03,kovtyukh06}.

\subsubsection{TYC 4144 329 2}

\begin{figure}
 \includegraphics[width=160mm]{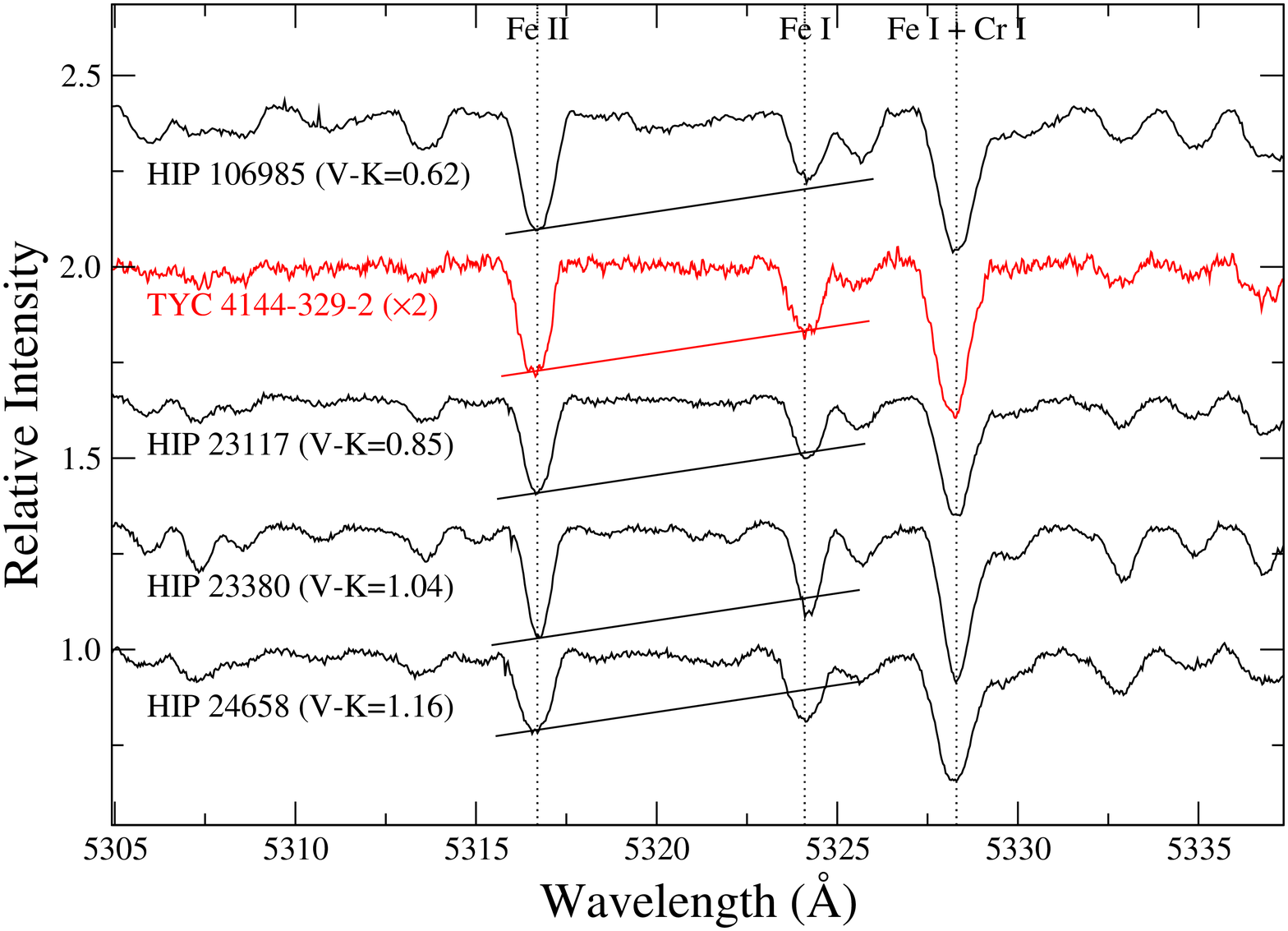}
\caption{\label{figsptype2}
         Keck HIRES spectra of TYC 4144 329 2 (two individual
         exposures averaged together) and main sequence, F-type stars not known
         to have any orbiting dust \citep[the spectrum of HIP 106985 comes from 
         the ELODIE archive;][]{moultaka04}. Comparison of temperature-sensitive
         absorption lines indicates that the temperature class of
         TYC 4144 329 2 is somewhere between an A7 star (V$-$K=0.62) and an
         F5 star (V$-$K=1.042). The slanted lines are added to help guide the
         eye and have their slopes fixed at the slope value determined from
         comparing the line strength ratios of the Fe~II and Fe~I absorption
         lines of TYC 4144 329 2. The best match appears to be around an F2
         star (V$-$K=0.85). Wavelengths in this figure and in Figure
         \ref{figacc} are plotted in the heliocentric vacuum scale.}
\end{figure}

An initial estimate of the temperature class of TYC 4144 329 2 came from a
comparison of our Keck HIRES spectra to similar resolution spectra of stars of
known spectral type that are not known to have any surrounding dust 
(Fig.\ \ref{figsptype2}). A rough comparison of different ions of the same
atom (Fe~I and Fe~II in the case of Fig.\ \ref{figsptype2})
allows us to determine that TYC 4144 329 2 has a temperature similar to that of
an F2-type star (T$_{\rm eff}$$\sim$7000 K). We attempted to further refine this
estimate with line depth ratios. Unfortunately, the broadened lines of TYC 4144 
329 2 complicate such diagnostics which typically rely on comparing a stronger 
absorption line with a weak absorption line. With a $v$sin$i$ of $\sim$30 \kms ,
the weak lines of TYC 4144 329 2 get subsumed into the continuum. As such the
most accurate line depth ratios are unavailable to this analysis, and the best 
we can do
is confirm our estimated temperature with lower limits. From comparing the
line depth ratios of Ti~I $\lambda$6126 to Si~I $\lambda$6155 with Fig.\ 2 of
\citet{kovtyukh03} we find that TYC 4144 329 2 must have a temperature greater
than $\sim$6500 K. For the line depth ratio of
V~I $\lambda$5727 to Si~I $\lambda$5772 we are able to barely detect both lines
after smoothing the spectrum with an 11-pixel boxcar,
although we do not feel the detection of V~I is greater than 3$\sigma$ 
significance. If we assume a detection then we get an effective temperature of
$\sim$6700 K when comparing to Fig.\ 2 of \citet{kovtyukh03}; if we take the
detection as an upper-limit the corresponding lower limit to the effective 
temperature of TYC 4144 329 2 is 6700 K. These two strong lower limits are
cosistent with our effective temperature estimate from comparison with
non-dusty field F-type stars. As such we feel confident that TYC 4144 329 2 has 
a temperature class of F2$\pm$2 subclasses.

\subsubsection{TYC 4144 329 1}

We make an initial estimate of the effective temperature of TYC 4144 329 1 from
broad band colors, finding a temperature class residing in the range of G8-K2
(the main uncertainty in this estimate is the luminosity class of 
TYC 4144 329 1).  With this estimate as a sanity check we then employed line 
depth ratios to refine the effective temperature measurement. Comparing the
residual intensity ratio of Fe~II $\lambda$6432 to Fe~I $\lambda$6430 against 
Fig.\ 2a of 
\citet{strassmeier90} yields a temperature class of G8$\pm$1 subclass if 
TYC 4144 329 1 is a giant or a subdwarf and K0$\pm$1 subclass if it is a dwarf.
We employ additionally the line depth ratios of Fe~I $\lambda$6241 to 
Si~I $\lambda$6244 and Fe~I $\lambda$6704 to Si~I $\lambda$6722 and find
temperatures of 4800$\pm$100 K for both when comparing against Fig.\ 1 of 
\citet{kovtyukh06}.

\begin{figure}
\begin{minipage}[t!]{85mm}
 \includegraphics[width=83mm]{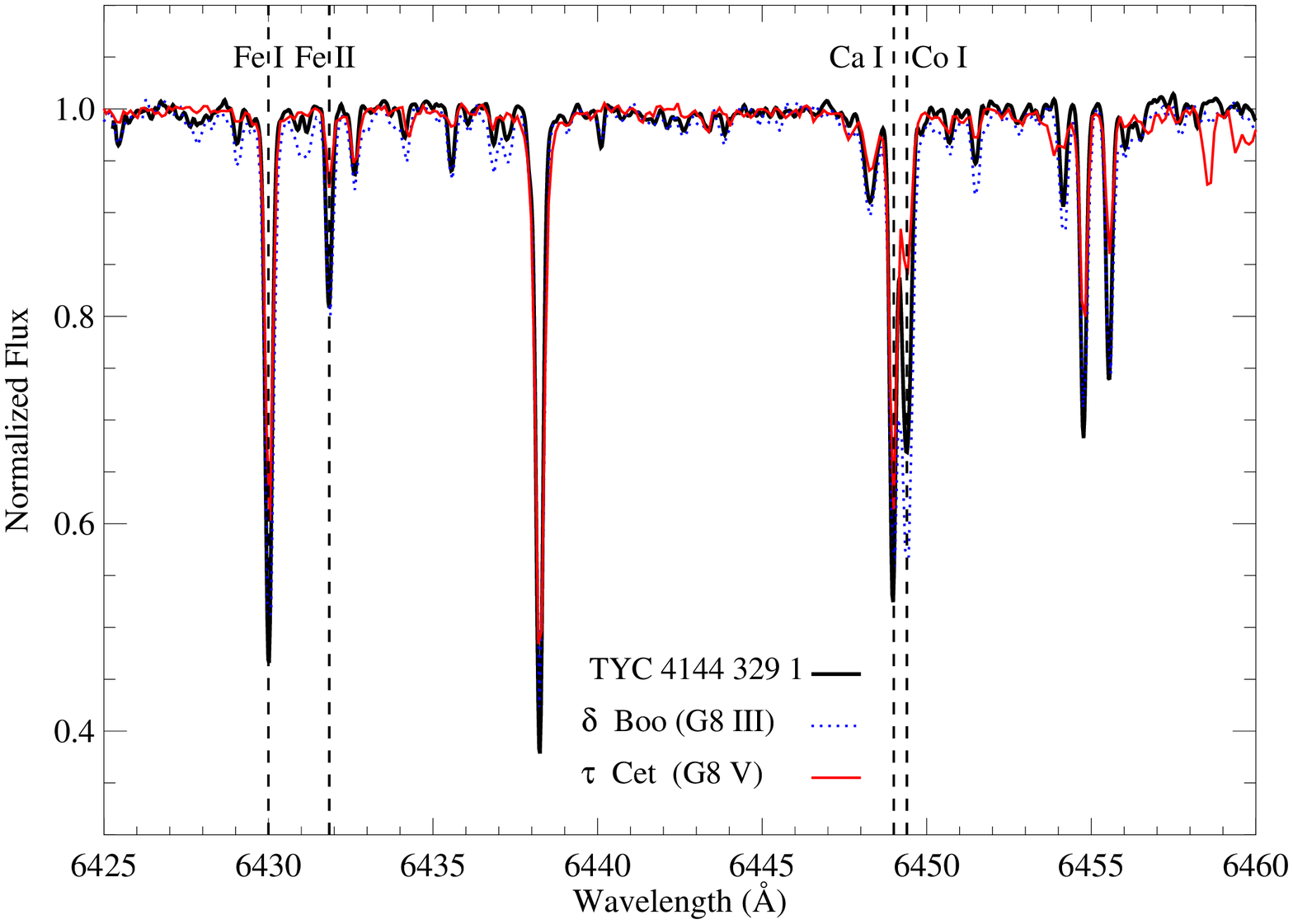}
\end{minipage}
\begin{minipage}[t!]{85mm}
 \includegraphics[width=83mm]{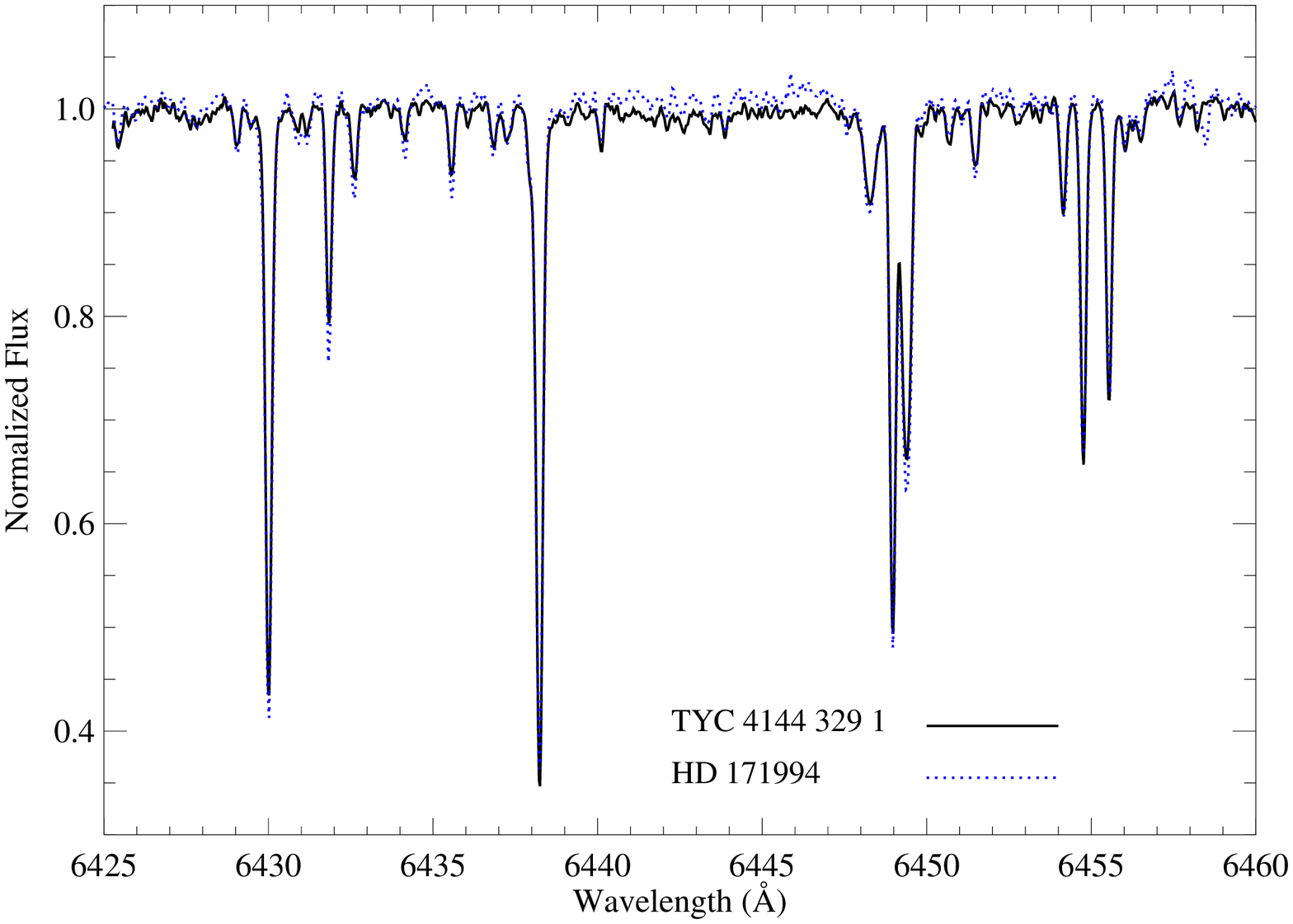}
\end{minipage}
\caption{\label{figsptype}
 {\it Both Panels:} Comparison of our HIRES spectrum of TYC 4144 329 1 with 
 $\delta$ Boo, $\tau$ Cet, and HD 171994 (G8~III, G8~V, and G8~IV
 respectively). All spectra presented in this figure have been shifted to the
 rest frame with wavelengths in air. {\it Left Panel:} Luminosity sensitive 
 features (like Co~I $\lambda$6450) indicate that TYC 4144 329 1 has
 a luminosity in between that of a giant star and a dwarf 
 star \citep{strassmeier90}. {\it Right Panel:} TYC 4144 329 1 with HD 171994 
 overplotted, both spectra have R$\sim$40,000. A
 S/N$\sim$190 HD 171994 spectrum was obtained from the ELODIE archive
 \citep{moultaka04}. We match our HIRES spectrum to the lower resolution pixel
 sampling of the ELODIE spectrum (pixel widths of $\sim$1.3 \kms\ and
 $\sim$3.1 \kms\ for HIRES and ELODIE, respectively). HD 171994 is our 
 qualitative best match to TYC 4144 329 1, and hence we adopt its metallicity 
 and absolute visual magnitude \citep{soubiran08} for TYC 4144 329 1. From the 
 adopted absolute visual magnitude we estimate a distance of $\sim$550 pc to the
 TYC 4144 329 system.}
\end{figure}

Our next step is to secure the luminosity class of TYC 4144 329 1 and finalize
its spectral type. As is illustrated in Fig.\ \ref{figsptype}, luminosity 
sensitive lines 
\citep[especially Co~I $\lambda$6450,][]{strassmeier90} show significant
deepening with increasing luminosity. Rough examination of Fig.\ \ref{figsptype}
shows that TYC 4144 329 1 has a luminosity intermediate to that of a dwarf and
a giant star. Comparing the residual intensity ratio of
Co~I $\lambda$6450 to Ca~I $\lambda$6449 with Figs.\ 2e and 3b from 
\citet{strassmeier90} shows good agreement with a subgiant classification. An
additional check of the luminosity class of TYC 4144 329 1 is obtained from
comparing the flux ratio of bandpasses around Fe~I $\lambda$4071 and 
Sr~II $\lambda$4077 to Fig.\ 1 of \citet{mamajek02}. For our determined 
effective temperature we find again that TYC 4144 329 1 has line ratios 
consistent with a subgiant. With all of the above diagnostics we are confident 
that the spectral type of TYC 4144 329 1 is G8~IV.

\begin{figure}
\centering
 \includegraphics[width=160mm]{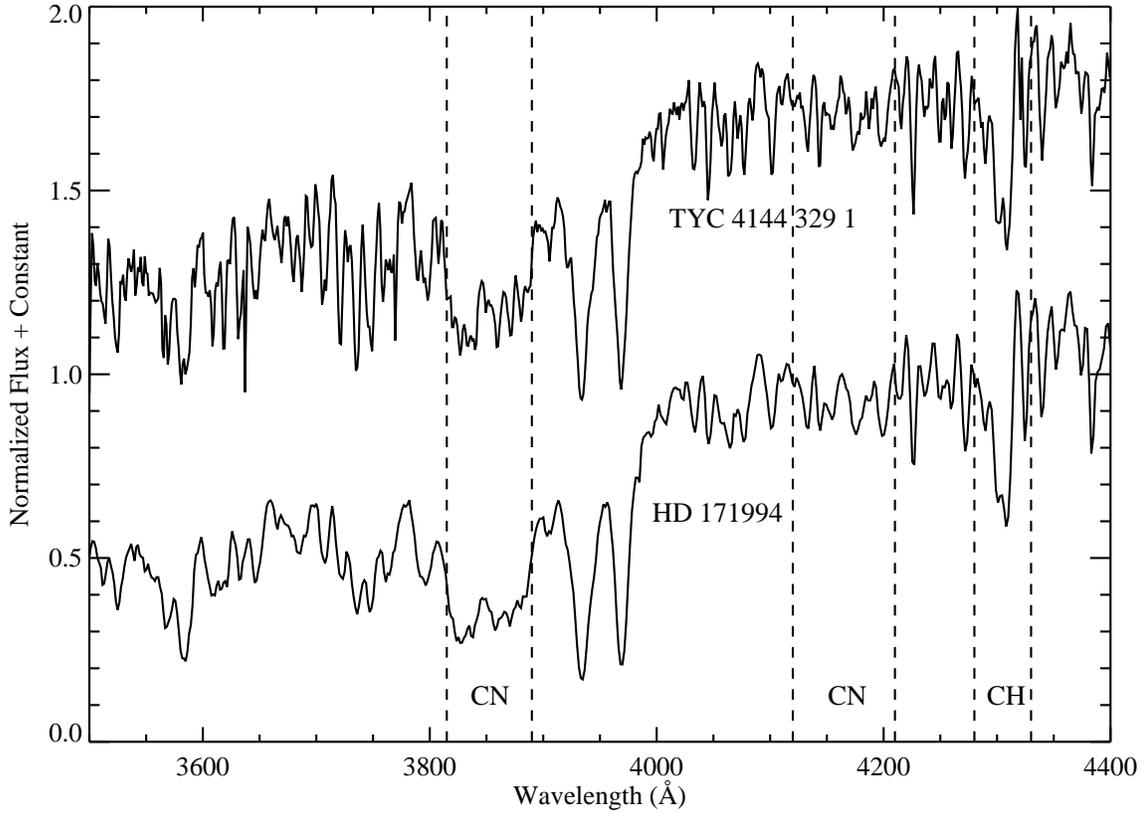}
\caption{\label{figkast}
 KAST blue side spectra of TYC 4144 329 1 and HD 171994 with spectral 
 resolutions of $\sim$3.4 and $\sim$6.5 \AA , respectively. Adverse conditions 
 present during the acquisition of each spectrum (see Section \ref{secobs} and 
 Table \ref{tabobs}) could
 have potentially affected the spectral calibrations. Although the spectral
 slopes are not identical, we find that line strengths, in particular the
 gravity sensitive CN and CH bands, are.}
\end{figure}

To complete the parameterization of TYC 4144 329 1 we sought robust estimates 
of its metallicity and luminosity. In lieu of attempting 
to model our HIRES and KAST data we decided to qualitatively match our 
spectra with similar resolution spectra of stars with known metallicities and
parallax measurements. Our best match to TYC 4144 329 1 is HD 171994 
(Figs.\ \ref{figsptype} and \ref{figkast}). As such we adopt parameters of 
HD 171994 \citep[see][]{soubiran08} for TYC 4144 329 1 (Table \ref{tabpars});
specifically the G8~IV spectral type classification, T$_{\rm eff}$, [Fe/H], 
and the luminosity of HD 171994. The Hipparcos measured
parallax of HD 171994 \citep[$\pi$=10.70$\pm$0.44 mas;][]{vanleeuwen07} 
enables us to estimate a distance to TYC 4144 329 1. Based on the adopted 
absolute visual magnitude of 1.46 we estimate TYC 4144 329 1 and TYC 4144 329 2 
(common space motions indicate the TYC 4144 329 system is bound, see 
Table \ref{tabpars}) reside at a distance of $\sim$550 pc.

\subsection{Excess Infrared Emission from TYC 4144 329 2}
\label{secirex}

TYC 4144 329 2 first came to our attention as part of a survey to identify 
TYCHO-2 main sequence stars with excess emission at mid- and far-infrared 
wavelengths (C. Melis {\it et al.} in preparation). To accomplish 
this we cross-correlated the TYCHO-2 catalog \citep{hog00} with the Infrared 
Astronomical Satellite catalogs \citep[IRAS,][]{moshir90}. The IRAS measurements
of the TYC 4144 329 
binary at 12, 25, and 60 \um\ (Fig.\ \ref{figirex}) indicate the presence of 
warm and cool 
dust around at least one of the two TYCHO-2 stars (the IRAS beam size is many 
times larger than the 5.5$\arcsec$ separation between the two stars). Subsequent
investigation of near-infrared fluxes at higher spatial resolution in the 
2MASS catalog \citep{cutri03} revealed a definite K$_{\rm s}$-band excess around
TYC 4144 329 2, the F2-type star, but not around the G8~IV optical primary. To 
ensure that the overall infrared excess was real and arising purely from 
TYC 4144 329 2 we observed the two stars with NIRC2. The relative L$\arcmin$ and
M$_{\rm s}$ fluxes indicate
that the near-infrared excess emission originates entirely from the F2 star. 
Hence, we attribute any longer wavelength excess emission (mid- and 
far-infrared) to the F2 star.

To estimate the dust temperature and the fraction of the stellar luminosity 
reradiated by the dust (L$_{\rm IR}$/L$_{\rm bol}$) we fit optical and 
near-infrared 
measurements out to J-band with a synthetic stellar atmosphere spectrum 
\citep{hau99} reddened by an interstellar extinction model (see caption to 
Fig.\ \ref{figirex}) along with two
blackbodies (having temperatures of 1200 and 120 K, see Fig.\ \ref{figirex}) 
that model the 
dust excess. The blackbody grains with a temperature of 120 K will reside at a
distance of $\sim$30 AU from TYC 4144 329 2.
This distance of 30 AU is derived under the assumption that TYC 4144 329 2 has 
the luminosity listed in Table \ref{tabiso}. Such a physical separation agrees 
well with 
the $\sim$45 AU distance to similar temperature dust orbiting the lithium-rich 
giant star HD 233517 \citep{jura03a}. Although the simple assumption of two 
narrow rings of dust 
provides a reasonable fit to the data, such a model is unlikely to be realistic 
considering that $\sim$17\% of the total luminosity of the star is absorbed and 
re-radiated at infrared wavelengths. To have so much light from the star 
processed by dust it is necessary to have an optically, vertically thick 
(e.g. flared) disk or a shell of dusty material surrounding the star.  





\subsection{Accretion Signatures}
\label{secacc}

\begin{figure}
\begin{minipage}[t!]{85mm}
 \includegraphics[width=83mm]{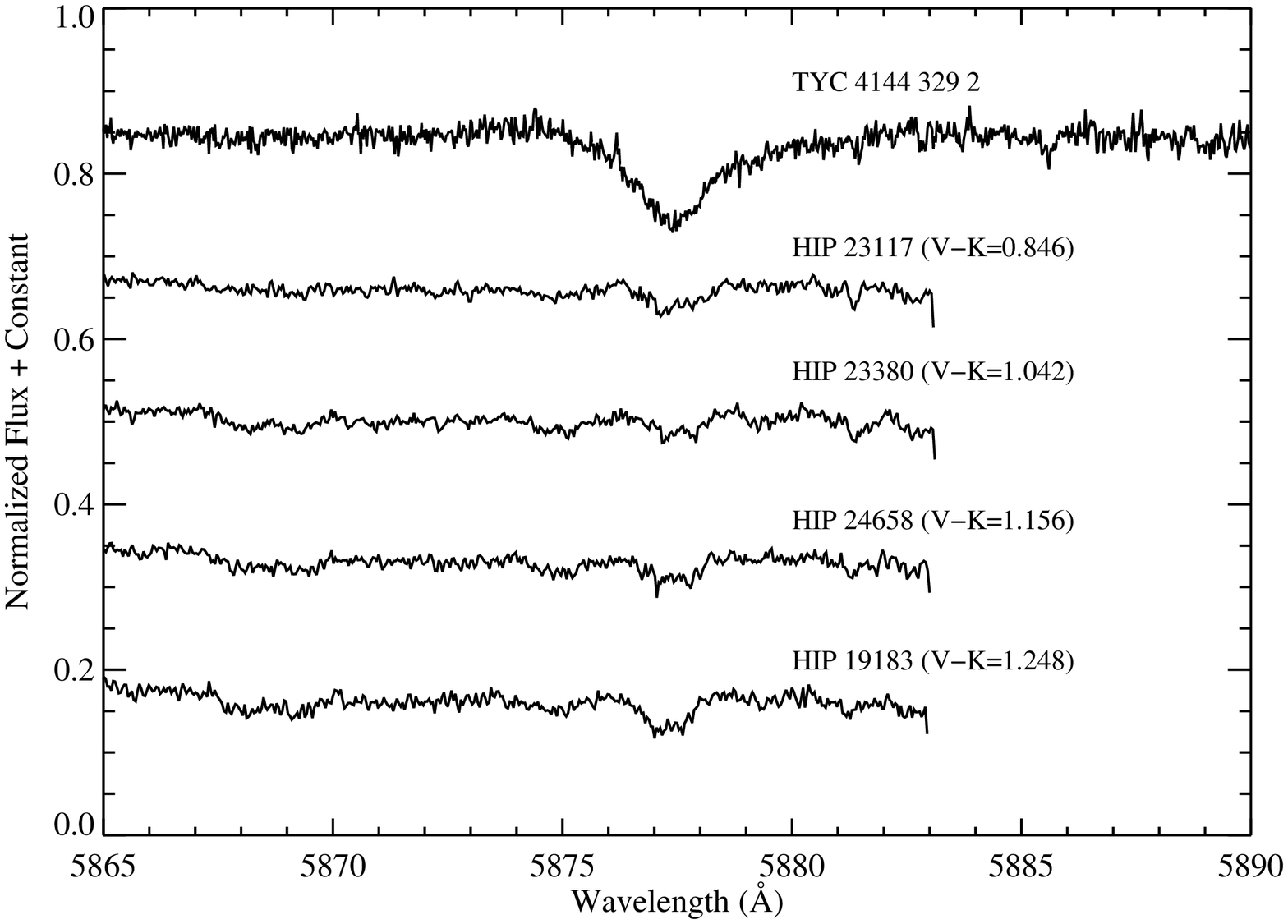}
\end{minipage}
\begin{minipage}[t!]{85mm}
 \includegraphics[width=83mm]{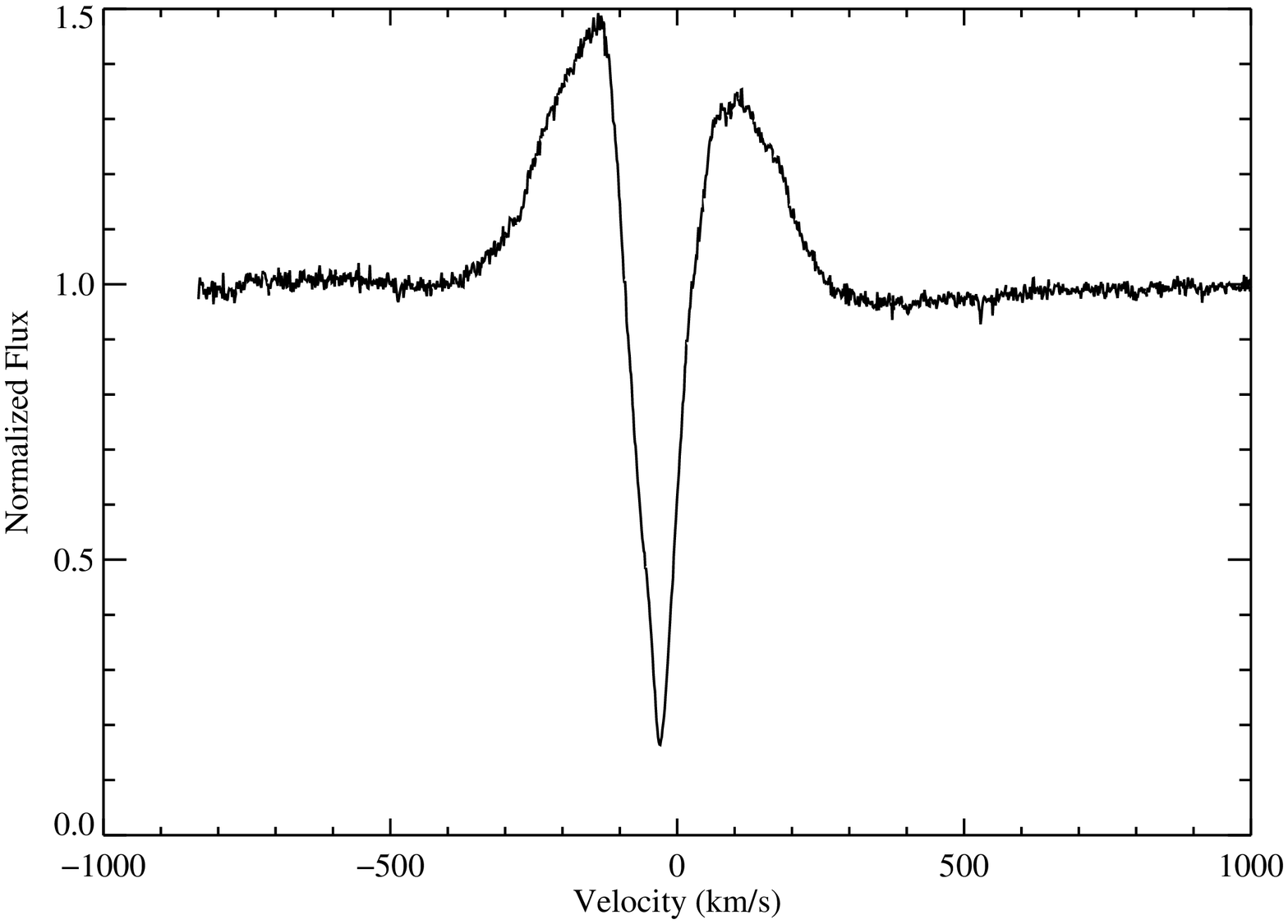}
\end{minipage}
\caption{\label{figacc}
         {\it Left Panel:} Helium $\lambda$5877 absorption feature in TYC 4144
         329 2 and non-dusty main sequence field F-type stars. These field
         F-type stars have He~I $\lambda$5877 EWs of $\sim$40 m\AA\ while TYC
         4144 329 2 has a He~I $\lambda$5877 EW of $\sim$290 m\AA . Such an
         enhanced He~I absorption feature could be the result of super-heated
         accreting material \citep{alencar00}. {\it Right
         Panel:} H$\alpha$ complex in TYC 4144 329 2. The broad velocity width
         of the emission feature is consistent with a star that is accreting
         gas \citep{white03}. The deep central absorption component might in
         part be due to disk gas orbiting the star. Similar features
         are seen in the H$\beta$ absorption line and the Na~D absorption
         lines (see Section \ref{secacc}).}
\end{figure}

The optical spectrum of TYC 4144 329 2 displays
an uncharacteristically broad and deep He~I $\lambda$5877 absorption line (Fig.\
\ref{figacc}). Although He~I $\lambda$5877 absorption lines are common in 
early- to mid-F type stars, they are typically fairly weak (Fig.\ \ref{figacc}).
The atmospheres of
young T Tauri stars, which are far too cool to excite this helium absorption
line, often show such a feature due to super-heated regions of accreting disk
gas \citep{alencar00}. Such a phenomenon could be boosting the strength of the
helium absorption line in TYC 4144 329 2. According to \citet{white03},
classical T Tauri stars with an H$\alpha$ emission line 10\% power width of 270
\kms\ or greater are accreting. The H$\alpha$ complex in TYC 4144 329 2 shows
both deep central absorption and double-peaked emission (Fig.\ \ref{figacc}). 
Such
double-peaked H$\alpha$ emission lines are common in T Tauri stars, who are
known to have substantial gaseous disks in orbit around them. Three methods
of measuring the H$\alpha$ emission-line width at 10\% of peak power
(from 10\% of the maximum flux in Fig.\ \ref{figacc}, from 10\% of the maximum 
of a
parabolic fit anchored at the two emission peaks, and from 10\% of a gaussian
anchored to the blue edge of the emission feature) yield an average of
556 \kms\ with a standard deviation of 30 \kms . Employing Eq. 1 from
\citet{natta04} we find a mass accretion rate from the H$\alpha$ 10\% width of
10$^{-7.5}$ M$_{\rm \odot}$ yr$^{-1}$. While characteristic of classical
T Tauri stars, no such rapid
accretion has ever been previously reported for a first-ascent
giant star. To the best of our knowledge there are no known published reports of
mass accretion at any rate, large or small, around such
stars\footnote{As mentioned in the Introduction, BP Psc may be a late-G
first-ascent giant star \citep{zuckerman08} with a mass accretion rate similar
to that of TYC 4144 329 2}. We estimate the accretion luminosity for TYC 4144
329 2 by comparing our determined accretion rate to young stars known
to be accreting at the same rate with measured accretion luminosities 
\citep{muzerolle98}. From this method we estimate the accretion luminosity for
TYC 4144 329 2 to be $\sim$0.1 L$_{\rm \odot}$, negligible compared to the 
stellar luminosity of TYC 4144 329 2.

There are shell-like absorption line components in the
Na~D, H$\alpha$, and H$\beta$ profiles. The deep H$\alpha$ absorption component
(Fig.\ \ref{figacc})
is blueshifted relative to the stellar velocity by $\sim$2.5 \kms , H$\beta$
has a similar velocity shift although there is unidentifiable structure
present that is blended with the line of interest preventing a velocity 
measurement. The narrow
components of the Na~D lines are blueshifted with respect to
the stellar velocity by $\sim$1.5 \kms\ and they are just resolved 
(FWHM$\sim$8 \kms ) by
HIRES (see Section \ref{secage}). Such characteristics could be indicative of
orbiting gas that is slowly moving away from the star. Similar
absorption features in the debris-disk system around $\beta$-Pictoris
were attributed to outer Na-rich regions of the disk,
although the Na lines in $\beta$-Pictoris are not shifted in radial
velocity relative to the star \citep{vidalmadjar86}. The narrow
component of the Na~D$_{2}$ line in $\beta$-Pictoris has an EW of 9.5$\pm$1
m\AA\ while TYC 4144 329 2 has a narrow Na~D$_{2}$ EW of 120$\pm$2 m\AA .

\subsection{Age of the TYC 4144 329 System}
\label{secage}

To aid in the determination of the age of the binary system we obtained and 
analyzed high-resolution
echelle spectra for both stars. From these optical spectra we estimate the age 
from the lithium content in the stellar photospheres, velocity widths of 
absorption lines, and Galactic space motion; details can be 
found in \citet{zuckerman04}. Of course, lithium content and velocity widths
are best used to tell if stars are particularly young. In the following 
analysis we only check that these parameters are consistent with our G8~IV 
spectral type of the optical primary. The lithium 6710 \AA\ absorption feature 
(wavelength in vacuum) is too weak to be detected in the spectrum of either 
star (Table \ref{tabpars}). Lithium content in a stellar atmosphere is mainly 
determined by the star's age and mass. These upper limits are consistent with 
roughly solar mass stars at the end
of their main sequence lifetimes. Similarly, the low measured $v$sin$i$ for TYC
4144 329 1, $\sim$3 \kms , is consistent with an evolved star
\citep{demedeiros95,graypalla89,schrijver93}. $v$sin$i$ was measured from the
full-width at half-maximum depth (FHWM) of single absorption lines in the
Keck HIRES spectra, which have intrinsic FWHM resolution of $\sim$8
\kms , a value we subtract in quadrature from the FWHM measured in the spectra.
A ROSAT All-Sky Survey non-detection of these stars is consistent with old
stellar ages \citep{zuckerman04}.

\begin{figure}
\centering
 \includegraphics[width=100mm]{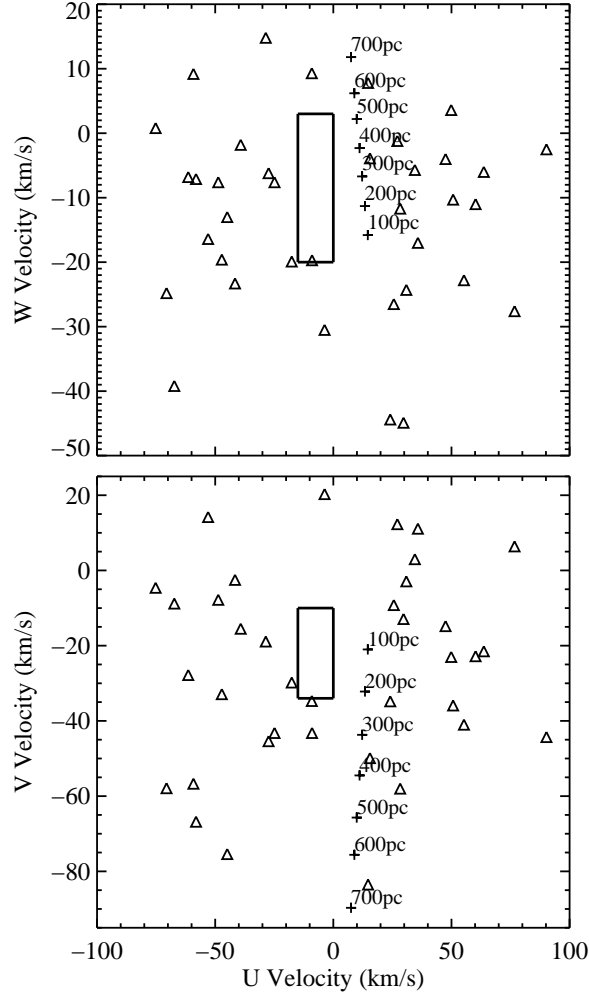}
\caption{\label{figuvw}
         Comparison of the UVW space motion for the TYC 4144 329 system 
         (crosses) with
         the UVW space motions of stars with known ages. We plot the entire 
         range of plausible distances for the TYC 4144 329 system as determined
         by ranging the luminosity of TYC 4144 329 1 from that of a main 
         sequence dwarf to that of a class III giant. Triangles in the plot
         are white dwarfs with a mean cooling age of 1.86 Gyr 
         \citep[where the cooling age does not include the main sequence 
         liftime of $\sim$3 Gyr for the typical white dwarf in this sample;][]{zuckerman03,bergeron01}. The rectangles
         represent the ``good'' UVW box for young stars as defined in 
         \citet{zuckerman04}. As stars age, the Galactic potential acts to 
         disperse them away from the young star boxes.}
\end{figure}

The above considerations suggest that the system is not young, in agreement 
with our spectral type determination of G8~IV for TYC 4144 329 1. As a further
refinement of our age estimate we can compare the space motions of the TYC 4144 
329 system with those of young stars and old white dwarfs (Fig.\ \ref{figuvw}). 
Velocities of the TYC 4144 329 system toward the center of our Milky Way galaxy,
around the Galactic Center, and perpendicular to the Galactic plane (U, V, W)
are calculated from TYCHO-2 proper motions, our estimated distance to the G8~IV 
star (see caption to Fig.\ \ref{figsptype} and discussion in 
Section \ref{secsptypes}), and the optical echelle measured 
radial velocities. As stars age, the Galactic potential acts to disperse them in
the U$-$V and U$-$W space. At a distance of $\sim$550 pc 
the TYC 4144 329 system comfortably lies within the typical space motions of old
white dwarf stars that have a mean cooling age of 1.86 Gyr
\citep[Fig.\ \ref{figuvw};][note that the cooling age does not include the
main-sequence lifetime, which is typically $\sim$3 Gyr for these white
dwarfs]{zuckerman03,bergeron01}.

\begin{figure}
 \includegraphics[width=160mm]{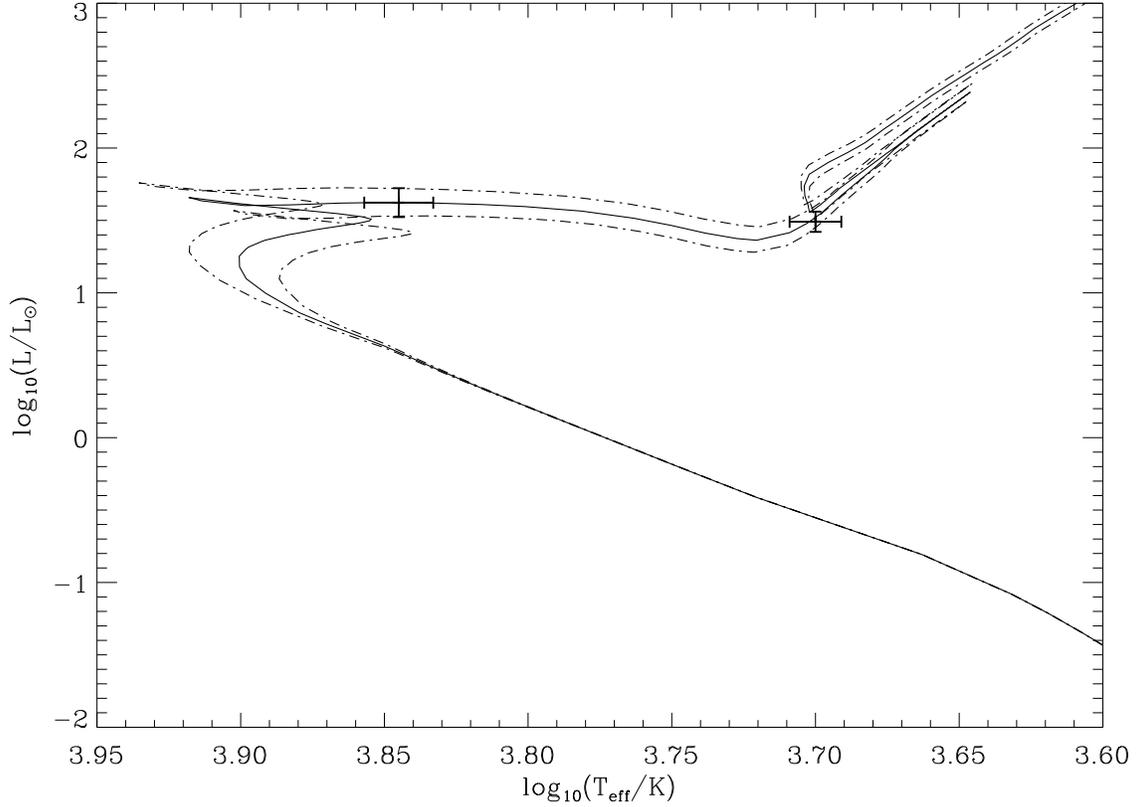}
\caption{\label{figiso} TYC 4144 329 system plotted on theoretical isochrones.
         Isochrones were obtained from the CMD 2.1 webform 
         (http://stev.oapd.inaf.it/cmd) and all have metallicity of 
         [Fe/H]=$-$0.2. The solid isochrone represents our best fit age for 
         the TYC 4144 329 system, 930 Myr. The two dot-dashed isochrones are
         for ages of 794 Myr and 1027 Myr. TYC 4144 329 1 is the point near 
         T$_{\rm eff}$ of 3.7 and L of 1.5 while TYC 4144 329 2 is the point
         near T$_{\rm eff}$ of 3.85 and L of 1.6. Exact values of effective
         temperature and luminosity and their associated errors can be found
         in Table \ref{tabiso}.}
\end{figure}

Finally, we determine more precisely the binary system's age by putting 
TYC 4144 329 1 and 2 on theoretical isochrones for evolved stars. We employed 
the CMD 2.1\footnote{http://stev.oapd.inaf.it/cmd} webform developed by 
L\'{e}o Girardi \citep{girardi00,marigo07,bertelli94,marigo08} to give model 
isochrones tailored to specific input metallicities. For our system metallicity 
of [Fe/H]$\sim$$-$0.2 we simultaneously fit the derived luminosities and 
effective temperatures for TYC 4144 329 1 and 2 to the model isochrone. We list 
in Table \ref{tabiso} the input parameters and ranges for these values that were
used for the model isochrone fit. From this fit we determine that the 
TYC 4144 329 system is 930$\pm$140 Myr old. We note that for
all acceptable ages obtained in this fashion TYC 4144 329 1 and TYC 4144 329 2
have evolved off the main sequence (see Fig.\ \ref{figiso}). 

Our above age determination suffers from a degeneracy between the size of the
dust grains and the luminosity of TYC 4144 329 2. In our analysis we have
assumed that the dust grains are interstellar-like as there is not yet any
evidence for grain growth beyond that size. It is plausible that the grains
in the disk surrounding TYC 4144 329 2 could be larger. Larger grains would
extinct the star more, and thus our derived luminosity would need to be
increased appropriately (see note to Table \ref{tabiso} for how we estimated
the luminosity of TYC 4144 329 2 assuming interstellar-like dust grains extinct
the star). However, there is a maximum grain size within the disk that is set
by the necessity that both TYC 4144 329 1 and 2 have the same age and thus can
be fit simultaneously to a single isochrone. To estimate this maximum grain size
we examined the range of TYC 4144 329 2 luminosities (with all other parameters 
held fixed) that still allow a simultaneous 
isochrone solution. The measured reddening, E$_{\rm B-V}$=0.32, translates to 
R=3.2 for the interstellar-like grain scenario (A$_{\rm V}$=1.0), where R is the
ratio of total to selective extinction. The maximum luminosity TYC 4144 329 2 
can have and still allow an isochrone solution, log(L$_2$/L$_{\rm \odot}$)=1.78,
corresponds to R=4.0 (A$_{\rm V}$=1.3), suggesting grains slightly larger than 
interstellar dust. If the dust grains are this large then the system could be as
young as 725 Myr old.




\section{Discussion}

Based on the disk constituents we postulate that the material accreting onto TYC
4144 329 2 originated from an object composed mostly of gas. The TYC 4144 329 
system resides at high galactic latitude and no known interstellar molecular 
clouds are nearby. Potential sources of large quantities of gas would include 
the atmosphere of TYC 4144 329 2 itself or a low-mass stellar or sub-stellar 
companion, possibly even a massive Jupiter from the star's own planetary system.
Main sequence stars with companions at small separations are known: the W UMa 
class of main sequence contact binaries and the ``hot Jupiter'' class of massive
planets. When the F2 star evolved off the main sequence there could have 
been interactions between it and its hypothetical short-orbital period companion
(second epoch radial velocity measurements show no evidence for a remnant 
short-orbital period companion around TYC 4144 329 2)
that could create a circumstellar disk. There are plausible models of the 
aftermath of the engulfment of a short-period companion by a first-ascent giant 
star that results in the formation of a disk \citep{jura03a,nordhaus06} as well 
as theoretical 
predictions of W UMa stars' endstates that involve the lower-mass companion 
transferring mass onto the evolved star and the formation of a circumbinary 
ring \citep{webbink03,shu79}.

Evidence in support of a companion ``consumption'' model comes from two
independent studies. The first is that of \citet{tokovinin06}, in which the 
frequency of tertiary companions to spectrosopic binaries with short periods
(P$<$30 days) is examined. They find that for very short orbital-period 
spectroscopic binaries (P$<$2.9 days) the incidence of tertiary components is
$\sim$100\% \citep[Fig.\ 14 of][]{tokovinin06}. Indeed, the hypothetical 
companion that was consumed by TYC 4144 329 2 must have been very short-period, 
as TYC 4144 329 2 has only barely passed beyond main-sequence core hydrogen 
burning.
The existence of
TYC 4144 329 1 \citep[whose separation of $\sim$3000 AU is consistent with the
separations of the tertiaries studied by][]{tokovinin06} is in accordance with
such a scenario. A second observational campaign includes recent work on binary 
post-AGB stars. It has been revealed that these
objects, when they host near-infrared excesses, are systematically found to be
associated with binarity \citep{deruyter06}. Confirmation that these 
near-infrared excesses are from Keplerian, circumbinary disks has come from 
interferometric observations of nearby systems 
\citep[][and references therein]{bujarrabal07,deroo07a,deroo07b}. In these
post-AGB systems the binarity is seen as a likely necessity for the formation
of a disk around the evolving giant star and its companion. The case for severe
binary interaction, the type of which we assume occurred for TYC 4144 329 2 but 
at a much earlier phase of its post-main sequence evolution, is quite strong for
these post-AGB stars \citep{deroo07b}.

We can make an estimate of the timescale of this phenomenon by noting that from 
of order 10$^{5}$ stars in our TYCHO-2/IRAS survey who could have shown 
L$_{\rm IR}$/L$_{\rm bol}$$\sim$17\%,
TYC 4144 329 2 is the only such detection. Assuming that this system is 
$\sim$1 Gyr old and that the phenomenon shows in one in 10$^{5}$ stars, we 
estimate that the phenomenon might last of order 10$^{4}$ years.

Lithium content in the F2 star's photosphere can be used to constrain the 
nature of the object that is the source of the accreting material. Brown 
dwarfs with mass $\lesssim$0.06 M$_{\rm \odot}$ should have a cosmic abundance 
of lithium as they never reach sufficient temperatures to burn it 
\citep{basri00}. A low-mass star, however, would have burned all of its lithium 
by the time it
is $\sim$1 Gyr old \citep{zuckerman04}. Lithium depletion in the atmosphere of
main sequence, early F-type stars is a slow process that becomes even more 
inefficient as the star ages and spins down \citep{talon98}. The timescale of 
substantial 
lithium depletion, on the order of several hundred Myr \citep{talon98}, is much
longer than our estimates of the timescale of the phenomenon occurring at TYC 
4144 329 2. Thus, if any brown dwarf with a cosmic abundance of lithium had been
accreted onto the photosphere of the F2 star we would 
definitely see a large equivalent-width lithium feature in our spectra. The fact
that TYC 4144 329 2 has begun to evolve off the main sequence does not affect
this statement, as for any reasonable initial mass of TYC 4144 329
2 a negligible amount of mass will be contained in the evolved star's slowly
rotating convective envelope \citep[see][Fig.\ 1e and Table 3]{mengel79}. Since 
we do not see such a lithium feature, and because we know that any accreted 
object must have a substantial gas mass, we can safely rule out any accreted 
companion with a mass $<$ 0.06 M$_{\rm \odot}$. It is also 
possible that the material in orbit around TYC 4144 329 2 originated from 
itself; for example, the star's atmosphere may have been ejected by interactions
with a common-envelope companion. In such a scenario it is not possible to 
constrain the mass of the companion.

\begin{figure}
 \includegraphics[width=160mm]{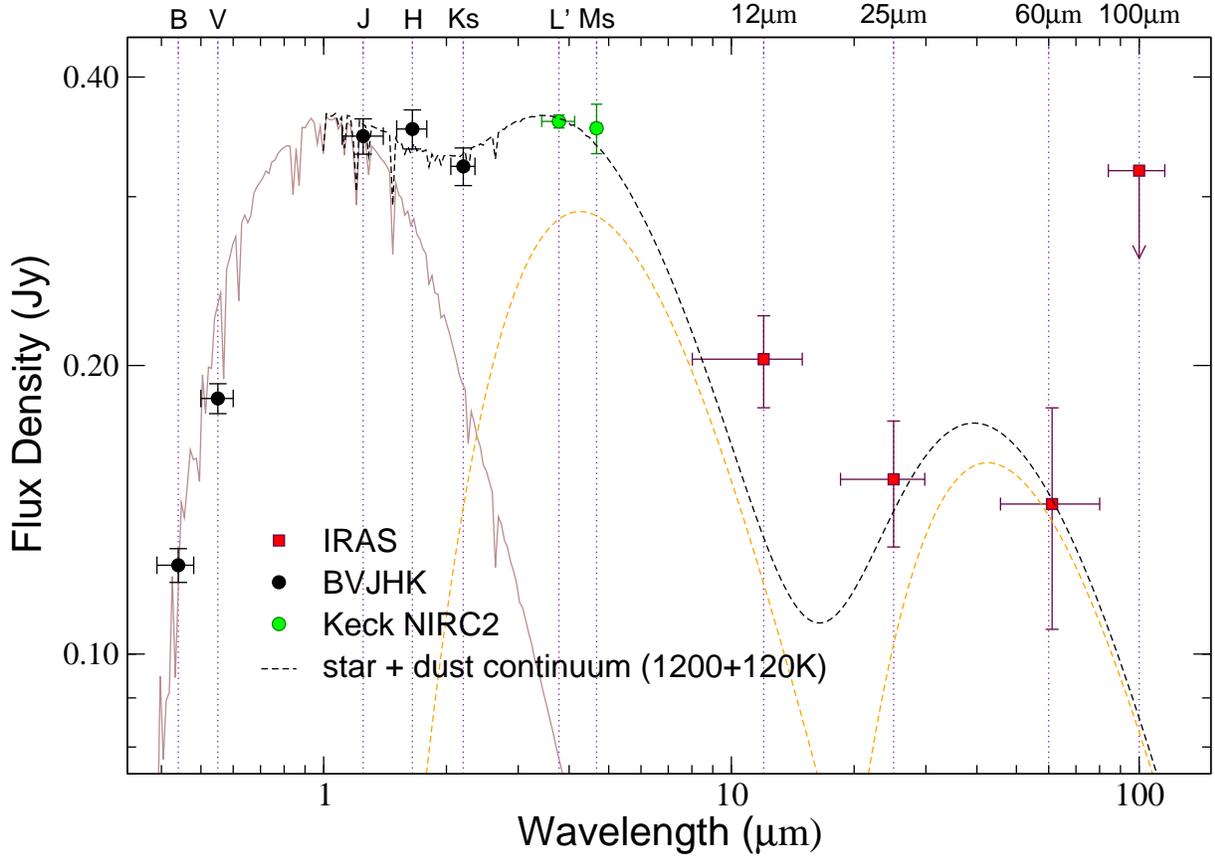}
\caption{\label{figirex}
         We detect significant infrared excesses from TYC 4144 329 2 at 
         wavelengths greater than $\sim$1 \um . The solid brown curve is a 
         synthetic stellar
         spectrum \citep{hau99} for a 7000 K effective temperature star reddened
         by an interstellar extinction model. The orange dashed curves 
         represent two dust continuum emission blackbodies fitted to dust 
         temperatures of 1200 and 120 K. The black dashed line, the
         sum of the above three curves, is a reasonable fit to most of the data
         points. The elevated position of the 12 \um\ IRAS point suggests either
         a silicate emission feature or some dust with temperature intermediate
         between 120 and 1200 K.  The total luminosity of the excess, determined
         by integrating under the data points between 1 \um\ and 100 \um , 
         is $\sim$17\% of  the luminosity of the star. The BV data points 
         (shortward of 1 \um ) are from the TYCHO-2 catalog, whereas the 
         JHK$_{\rm s}$ data points (longward of 1 \um ) are from the 2MASS 
         database.  The green data points are our own measurements 
         obtained on UT 23 June 2007 with the Keck NIRC2 camera. The L$\arcmin$ 
         flux error bar is smaller than the point size on the plot. The 
         red data points are from the IRAS Faint Source Catalog. The horizontal
         bars indicate the filter bandwidths.} 
\end{figure}

In closing we note one final implication of the nature of the material 
surrounding TYC 4144 329 2. For such optically thick material, a flat dust disk 
with an inner radius of order 30 stellar radii (determined from the hot 
component of the spectral energy distribution, Fig.\ \ref{figirex}) cannot 
absorb 17\% of TYC 4144 329 2's luminosity \citep{jura03b}. A disk must be 
puffed up or warped \citep{pringle96} to have the right geometry to process this
amount of radiation. It is perhaps possible that a remnant gas-giant planet 
survived the event that occurred at TYC 4144 329 2 and its orbit through the 
dusty and gaseous disk gravitationally perturbs the disk material. In addition, 
a current generation of planetesimals may be forming in the gaseous and dusty 
material as is expected for similar disks found around T Tauri stars. If this 
were the case, then hypothetical in-situ radioactive dating of rocky objects 
around TYC 4144 329 2 at some point in the future could reveal two significantly
different ages, corresponding to
the two epochs of planet formation that may have occurred in this system. Such
a situation would be markedly different from that in the present-day Solar
system, where the bulk of the asteroidal material is thought to have formed
approximately contemporaneously with the Sun.

\acknowledgments

We are indebted to Frank Fekel for providing advice and comparison spectra to 
aid our determination of the spectral type of TYC 4144 329 1. We thank Jason X. 
Prochaska and Michael Murphy for obtaining the UT March 11 2008 KAST spectra for
us. We thank Bruce Draine for kindly providing us with reddening models tailored
to our data. We thank Bruce Macintosh and Christian Marois for obtaining one 
epoch of NIRC2 AO data for us. We are grateful to Peter Eggleton for helpful 
advice. Support for S.A.M. was provided by NASA through the Spitzer Fellowship 
Program, under award 1273192. The data presented herein were obtained at the 
W.M. Keck Observatory, which is operated as a scientific partnership among the 
California Institute of Technology, the University of California and the 
National Aeronautics and Space Administration. The Observatory was made possible
by the generous financial support of the W.M. Keck Foundation. This work was 
partially supported by the NASA ADP grant ADP06-0095 and by a NASA grant to 
UCLA. This research has made use of the VizieR service and of data products from
the Two Micron All Sky Survey.



{\it Facilities:} \facility{IRAS ()}, \facility{Keck:I (HIRES)}, \facility{Keck:II (NIRC2-NGSAO)}, \facility{Shane (KAST)}

\bibliographystyle{natbib}
\bibliography{archive}

\clearpage





\clearpage

\clearpage

\clearpage

\begin{deluxetable}{lccccccc}
\rotate
\tablecaption{Observations Summary \label{tabobs}}
\tablewidth{0pt}
\tablehead{
\colhead{UT Date} & \colhead{Instrument} & \colhead{Conditions} & \colhead{Setup} & \colhead{Coverage} & \colhead{Resolution\tablenotemark{a}} & \colhead{S/N\tablenotemark{b}} & \colhead{$\lambda$ of S/N\tablenotemark{c}}
}
\startdata
23 Jun 2007 & NIRC2 & Clear & 0.01$\arcsec$/pixel & JHK$_{\rm s}$L$\arcmin$M$_{\rm s}$ & $-$ & $-$ & $-$ \\
25 Oct 2007 & NIRC2 & Non-Photometric & 0.01$\arcsec$/pixel & JHK$_{\rm s}$L$\arcmin$ & $-$ & $-$ & $-$ \\
\multirow{2}{*}{05 May 2007} & \multirow{2}{*}{HIRES} & \multirow{2}{*}{Clear} & UV Collimator - Blue & 3100-5900 \AA\ & 40700 & 100/100 & 5850 \AA\ \\
                             &                        &                        & UV Collimator - Red & 4100-7000 \AA\ & 39600 & 100/100 & 6700 \AA\ \\
14 Feb 2008 & HIRES & Cloudy & UV Collimator & 3000-5900 \AA\ & 39700 & 80/50 & 5500 \AA\ \\
\multirow{2}{*}{11 Mar 2008\tablenotemark{d}} & \multirow{2}{*}{KAST} & \multirow{2}{*}{Cloudy} & 830/3460 & 3000-4550 \AA\ & $\sim$3.4 \AA\ & 35/35 & 4000 \AA\ \\
                                       &                       &                         & 600/7500 & 4500-7000 \AA\ & $\sim$4.5 \AA\ & 100/100 & 6000 \AA\ \\
\multirow{2}{*}{28 Jun 2008\tablenotemark{e}} & \multirow{2}{*}{KAST} & \multirow{2}{*}{See text} & 600/4310 & 3300-5500 \AA\ & $\sim$6.5 \AA\ & 75 & 4800 \AA\ \\
                                       &                       &                           & 830/8460 & 5500-7300 \AA\ & $\sim$3.4 \AA\ & 75 & 6000 \AA\ \\
\enddata
\tablenotetext{a}{Resolutions were measured from the FWHM of single arclines in our comparison spectra.}
\tablenotetext{b}{S/N for TYC 4144 329 1/TYC 4144 329 2 unless otherwise noted.}
\tablenotetext{c}{Wavelength where S/N measurement was made in the spectrum.}
\tablenotetext{d}{Observations used the 1.0$\arcsec$ slit and the d46 dichoric.}
\tablenotetext{e}{Observations used the 1.0$\arcsec$ slit and the d55 dichoric. This entry is for the comparison star, HD 171994.}
\end{deluxetable}

\clearpage

\begin{table}
\caption{Parameters of the TYC 4144 329 binary star system \label{tabpars}}
\centering
\begin{tabular}{ccc} 
\tableline
\multicolumn{3}{c}{Individual Stars} \\
\tableline
\tableline
 & TYC 4144 329 1 & TYC 4144 329 2 \\
\tableline
RA (J2000) & 10 23 10.57 & 10 23 09.88 \\
DEC (J2000) & +61 36 46.0 & +61 36 43.5 \\
Sp. Type & G8~IV & F2 \\
Vmag & 10.12 & 10.75 \\
T$_{\rm eff}$ (K) & 5014 & 7000$\pm$200 \\
pmRA (mas yr$^{-1}$) & $-$1.7 $\pm$ 2.8 & +2.4 $\pm$ 2.9 \\
pmDE (mas yr$^{-1}$) & $-$25.5 $\pm$ 2.8 & $-$24.3 $\pm$ 2.8 \\
RV (\kms ) & $-$27.48 $\pm$ 0.18 & $-$27.52 $\pm$ 1.35 \\
Li~I 6710 \AA\ EW (m\AA) & $< 2$ & $< 20$ \\
$v$sin$i$ (\kms ) & 3 & 31 \\
\tableline
\multicolumn{3}{c}{System} \\
\tableline
\tableline
Distance from Earth & \multicolumn{2}{c}{$\sim$550 pc} \\
Separation Between Stars & \multicolumn{2}{c}{5.5$\arcsec$ (projected separation of 3025 AU)} \\
UVW Space Motions (\kms ) & \multicolumn{2}{c}{+9.4, $-$71.3, +4.4} \\
{\rm [}Fe/H] & \multicolumn{2}{c}{$-$0.2} \\
Age & \multicolumn{2}{c}{930$\pm$140 Myr} \\
\tableline
\end{tabular}
\tablecomments{See the text for discussion of stellar parameters. 
\citet{samus04} detect 
variability in visual combined light measurements of this system. They report 
maximum and minimum V-band magnitudes of 9.8 and 10.6, respectively. Similarly, 
combined light measurements of this system in the TASS Mark~IV catalog 
\citep{droege06} show significant changes in brightness indicating that one or 
both of the stars are variable. Two epochs of our NIRC2 data (UT 23 June 2007 
and UT 25 October 2007) combined with 2MASS measurements (UT 14 November 
1999) indicate variability for TYC 4144 329 2 in the J-band at the 
tenth-of-a-magnitude level, but no such variability at longer wavelengths 
(H, K$_{\rm s}$, or L$\arcmin$).}

\end{table}

\clearpage

\begin{table}
\caption{Model Isochrone Fitting Parameters \label{tabiso}}
\centering
\begin{tabular}{ccc}
\tableline
                     &  TYC 4144 329 1  &  TYC 4144 329 2 \\
\tableline \tableline
\multicolumn{3}{c}{Inputs} \\
\tableline
log(L/L$_{\rm \odot}$) &   1.49$\pm$0.07  & 1.62$\pm$0.10 \\
log(T$_{\rm eff}$/K) &  3.700$\pm$0.009   & 3.845$\pm$0.012 \\
\tableline
\multicolumn{3}{c}{Outputs} \\
\tableline
M$_{\rm *}$ (M$_{\rm \odot}$) & 2.07$\pm$0.10 & 2.06$\pm$0.11 \\
log(g) &  3.04$\pm$0.11     &  3.46$\pm$0.08 \\
Age (Myr) & \multicolumn{2}{c}{930$\pm$140} \\
\tableline
\end{tabular}
\tablecomments{The range in log(T$_{\rm eff}$/K) corresponds to roughly 
$\pm$100 K for TYC 4144 329 1 and $\pm$200 K for TYC 4144 329 2. We determine 
the range in log(L/L$_{\rm \odot}$) for TYC 4144 329 1
from photometric and temperature errors that are propogated into its luminosity.
We adopt a wider range in log(L/L$_{\rm \odot}$) for TYC 4144 329 2 to 
accomodate the contributions from variability and accretion luminosity.
We determined the luminosity of TYC 4144 329 2 by matching the monochromatic
J-band magnitude to a blackbody with effective temperature of 7000 K at a 
distance of 550 pc. We then increased this resulting luminosity by 17\% to 
account for stellar radiation re-processed into the infrared (see 
Fig.\ \ref{figirex}) and again by a factor of 1.28 to compensate for exctinction
assuming there were interstellar medium-like grains in TYC 4144 329 2's dusty 
disk. Masses and
surface gravities were calculated by averaging the maximum and minimum values 
obtained from all solutions to the isochrone fits (the isochrone fit output for
the surface gravity of TYC 4144 329 1 is consistent with HD 171994's surface 
gravity from \citet{soubiran08}, where HD 171994 is the comparison star from 
which we adopted the TYC 4144 329 1 input parameters).}

\end{table}






\end{document}